# Towards the relationship between AIGC in manuscript writing and author profiles: evidence from preprints in LLMs

Jialin Liu and Yi Bu

*{liu_jialin, buyi}@pku.edu.cn*
0000-0003-4347-6458, 0000-0003-2549-4580
Department of Information Management, Peking University, China

**Abstract**
AIGC tools such as ChatGPT have profoundly changed scientific research, leading to widespread attention on its use on academic writing. Leveraging preprints from large language models, this study examined the use of AIGC in manuscript writing and its correlation with author profiles. We found that, (1) since the release of ChatGPT, the likelihood of abstracts being AI-generated has gradually increased; (2) scientists from English-speaking countries are less likely to use AIGC tools for writing assistance, while those from countries with linguistic differences from English are more likely to use these tools; (3) there is weak correlation between a paper's AI-generated probability and authors' academic performance; and (4) authors who have previously published papers with high AI-generated probabilities are more likely to continue using AIGC tools. We believe that this paper provides insightful results for relevant policies and norms and in enhancing the understanding of the relationship between humans and AI.

## 1. Introduction

Launched on November 30, 2022, ChatGPT is a powerful chatbot based on large language models (LLMs) developed by OpenAI. It has revolutionized many people's perceptions of AI, making a fundamental influence on both daily life and scientific research. Undoubtedly, LLMs was one of the hottest research fields in 2023. In the field of AI-content detection[1], scientific publications have become a popular research subject. Many works in AI-content detection have leveraged texts from publications to build benchmark datasets (e.g., Mosca et al., 2023; Yu et al., 2023) or train their prediction models (e.g., Liu et al., 2023). The adoption of artificial intelligence-generated content (AIGC) in manuscript writing has sparked discussions (Kendall & da Silva, 2024; Prillaman, 2024; Stokel-Walker, 2023) about its suitability for manuscript writing and whether ChatGPT should be considered an author. Recently, utilizing online tools or prediction models, several studies have quantified the use of AIGC in scientific publications (Akram, 2024; Liang, Zhang, et al., 2024; Picazo-Sanchez & Ortiz-Martin, 2024) and peer review comments (Chawla, 2024; Liang, Izzo, et al., 2024). These studies provide a global overview of AIGC adoption in academic writing and identify certain words disproportionately appeared in AI-generated text.

However, most previous quantitative studies have merely focused on predicting generative AI outputs and lacked in-depth analysis on the underlying factors of use of AIGC. We believe that it is crucial to better understand the interactions between AIGC writing assistance tools and scientists, and more importantly, AI and human. This paper analysed the abstracts of recent publications and examines their AI-generated probabilities by leveraging two research questions: **(1)** To what extent do scientists utilize AIGC for manuscript writing, and how has this evolved since the release of ChatGPT? **(2)** How is the adoption of AIGC in academic writing related to authors' profiles, including their native languages, academic performance, and history of AIGC use?

---
[1] https://github.com/ICTMCG/Awesome-Machine-Generated-Text?tab=readme-ov-file#human-detection

## 2. Methodology
### 2.1. Dataset
We utilized metadata from scientific publications posted on the arXiv.org website. ArXiv serves as an open-access repository of preprints in STEM fields. We downloaded the arXiv dataset from Kaggle[2] on March 12, 2024, including essential bibliographic details such as titles, authors, abstracts, subject categories, and update histories for over 1.7 million papers. In the current study, we selected a subset of papers in the area of LLMs, a field where researchers are typically more familiar with and have better access to AIGC tools. Specifically, we selected papers within the computer science category whose title contained keywords (with some pre-processing) related to generative AI, including "AI generated", "automatically-generated", "computer-generated", "content generation", "generated text", "generative AI", "generative model", "gpt", "large language model", "llm", "machine generated", and "text generation". To analyse the AI-generated probability, we only included papers firstly submitted in 2023 and excluded those abstracts with fewer than 100 words, resulting in the final dataset of 4,889 papers. The monthly distribution of these papers is shown in Figures 1a and 1b.

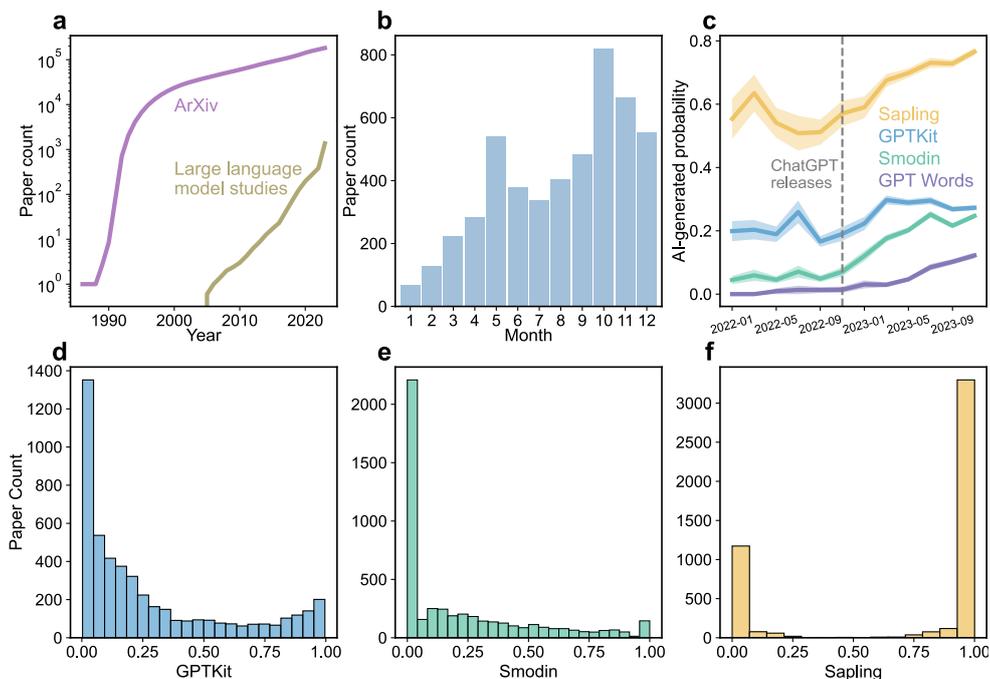

Figure 1. (a) Yearly distribution of papers submitted to arXiv in 1986-2023. (b) Monthly distribution of papers on LLMs in 2023. (c) Dynamics of the average AI-generated probabilities measured by four methods, spanning from January 2022 to December 2023; the enveloped areas represent 95% confidence intervals. (d-f) Distribution of AI-generated probabilities as predicted by three online detection platforms: GPTKit (d), Smodin (e), and Sapling (f).

### 2.2. AI-generated probability
We used abstracts as a proximity of authors' AI usage in their publications instead of full text. To quantify the AI-generated probability, we employed four distinct methodologies, using three existing online AI-content detection tools: GPTKit [3], Smodin [4], and Sapling [5]. By invoking the APIs of these platforms, we obtained the likelihood of abstracts being AI-generated, with output indicators ranging between 0 and 1. The detections always yielded two kinds of probabilities, one for the entire abstract and the other for individual sentences. Given that these two probabilities are based upon different models, it is unfair to weigh the separated probabilities together. Therefore, we used the overall prediction score as the AI-generated probability in the subsequent analyses. We also dichotomized the probability at a threshold of 0.9, grouped paper into high AI-generated abstracts and other abstracts, and repeated the analysis, and the results were robust. Our fourth AI-generated text detection method directly reviewed the abstracts without any models, according to whether the text contained any "GPT words". Recent works have found that there are some words that generative AI commonly used in scientific papers such as "pivotal", "intricate", "realm", "showcasing" (Liang, Zhang, et al., 2024), and "delve"[6]. We assigned an AI-generated probability of 1 if an abstract contains any of these five words, otherwise, we assigned a value of 0.

---
[2] https://www.kaggle.com/datasets/Cornell-University/arxiv

*2.3 Author-Level Information*

We detected native languages of authors via the "nationalize service" [7]. This tool infers nationality from authors' names and returns a list of countries in order of likelihood. If a nationality cannot be inferred from the name, the result will be "unknown". We considered the country with the highest probability to be the author's nationality, and those papers (8.3% of our dataset) with any author whose nationality cannot be detected through their names were excluded from following analyses. To access the native languages for authors from certain countries, we referenced the Geodist dataset (Mayer & Zignago, 2011), which lists the official languages spoken by over 20% of the population in each country.

As arXiv did not provide author name disambiguation, we extracted author academic performance data from Google Scholar, including citation counts, citation counts from the last five years, *h*-index, and *h*-index from the last five years. Specifically, we firstly used the Scholarly Python package (Cholewiak et al., 2021) on April 1, 2024 to retrace the focused publications by searching for the first author's name and title, allowing us to identify authors' unique author identifiers for those who have verified their Google Scholar profiles. Then we used the Scholarly to search the author with the author ID to get the required four indicators. Yet, due to the limitations of web scraping, which only allows retrieval of author lists displayed on Google Scholar search pages, the author lists of articles with a large number of authors are likely to be incomplete. Therefore, we only considered the academic performance of the first author, and papers (19.6% of our dataset) whose first author lacked a Google Scholar profile were also excluded from further performance analysis.

**3. Results**

*3.1 AI-generated probability increases gradually after ChatGPT release*

Figures 1d-1f show the distribution of the AI-generated probabilities predicted by three online tools. We observed that the variable distribution predicted by GPTKit and Smodin are similar. While the AI-generated probability for most abstracts is very low and the number of papers decreases as the probability increases, a small proportion of papers (7.3% for GPTKIT and 3.7% for Smodin) exhibit an AI-generated probability greater than 0.9. This aligns with

---

[3] https://gptkit.ai/
[4] https://smodin.io/
[5] https://sapling.ai/
[6] https://pshapira.net/2024/03/31/delving-into-delve/
[7] https://nationalize.io/

predictions based on certain "GPT words," indicating that 8.4% of papers in our dataset contain words frequently used by AIGC tools. However, Sapling appears to focus more on the "recall rate" of AI-generated text, with 68.5% of abstracts having a GPT probability above 0.9. It is important to clarify that this high proportion does not necessarily imply that these abstracts were definitely AI-generated. We conducted correlation analysis and found that the correlation among our four measurements is relatively low, indicating that the models behind these platforms are quite different and that they capture different facets of AI-generated text. In Figure 1c, we plotted the average AI-generated probabilities of preprints firstly submitted between January 2022 and December 2023. We found that, with the exception of GPTKit, there is an apparent increase in the probability of AI-generated text at the beginning of 2023, and the figure grows steadily throughout 2023. The trend reflects the shift of writing style after the release of ChatGPT and the influence of AIGC on abstract writing.

*3.2 Non-native English speakers are more likely to adopt AIGC tools in academic writing*
To explore the relationship between authors' native languages and the use of AIGC tools in scientific publications, we divided the papers into two groups based on whether at least one author is predicted to come from a country where English is the official language. Figure 2a displays the average of AI-generated probability from the four measurements. We employed the Mann-Whitney U-test to compare the corresponding probabilities between the two groups, revealing significant differences. For all four metrics, abstracts authored by individuals from non-English-speaking countries showed a higher probability of being AI-generated. Furthermore, we made the comparisons among seven non-English sub-groups with the greatest number of papers. Results from the Kruskal-Wallis H-test indicated significant differences among these seven groups. Compared to authors speaking French, German, and Hindi, teams with authors speaking Chinese, Italian, Japanese, and Korean tend to produce abstracts with higher AI-generated probabilities (Figures 2c-2f). These findings suggest that linguistic habits are closely linked to the use of AIGC, and scientists from countries with weaker English environments are more likely to utilize AIGC tools in academic writing.

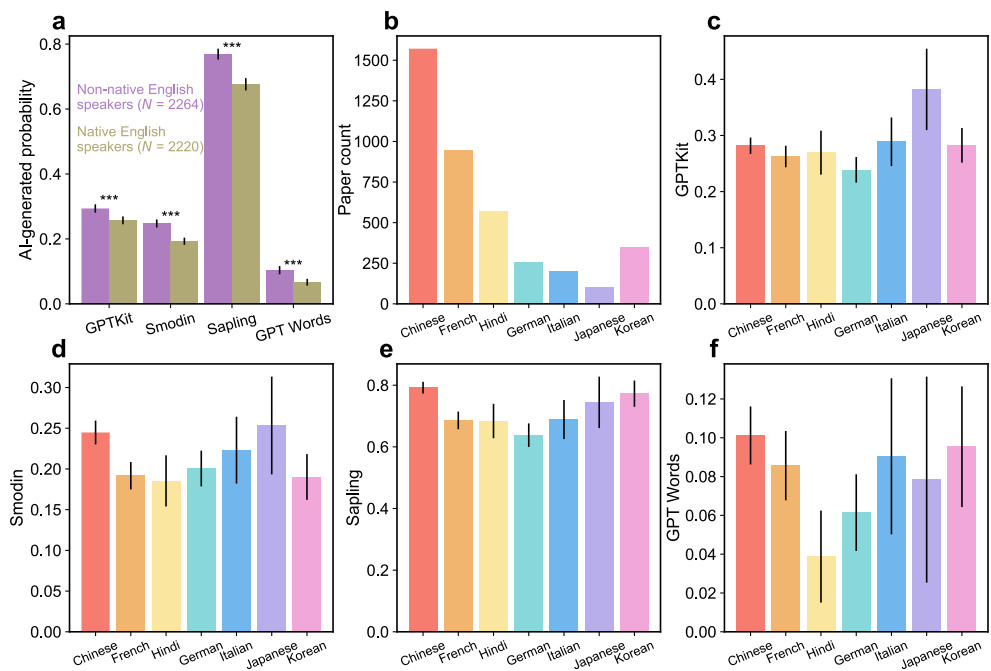

Figure 2. (a) Comparison of AI-generated probabilities between non-native English speakers and native English speakers, with error bars representing 95% confidence intervals. Significance levels are indicated as follows: *$p < 0.05$, ** $p < 0.01$, *** $p < 0.001$; the same

for Figure 3 and Table 1. (b) Number of papers authored by at least one author from seven countries where English is not the official language. (c-f) Average AI-generated probabilities for non-native English speakers versus native English speakers as predicted by GPTKit (c), Smodin (d), Sapling (e), and "GPT words" (f).

*3.3 Correlations between academic performance and adoption of AIGC are weak*

In terms of the academic performance of authors, we similarly divided the papers into two groups and compared the AI-generated probability based on whether the first author's citation counts, citation counts from the last five years, $h$-index, and $h$-index from the last five years ranked within the top 5% of our dataset. Yet, as shown in Figures 3a-3d, the findings based upon different measurements were inconsistent, and most of group differences were insignificant, which indicates the low correlation. Additionally, we calculated the Spearman correlation coefficient between the AI-generated probabilities and academic performance without grouping. As we can observe in Table 1, the correlation coefficients are very small and statistically insignificant. A possible explanation is that the academic performance of the first author does not precisely reflect the achievements of the entire team, and the abstract could be written by any team member. To mitigate this bias, we re-analysed the data across different team sizes (from 2 to 10), but finally yielding similar results. These findings suggest that the correlation is insignificant even in small teams, which further support our previous findings. Consequently, we concluded that there is no significant relationship between AI-generated probabilities and the authors' academic performance.

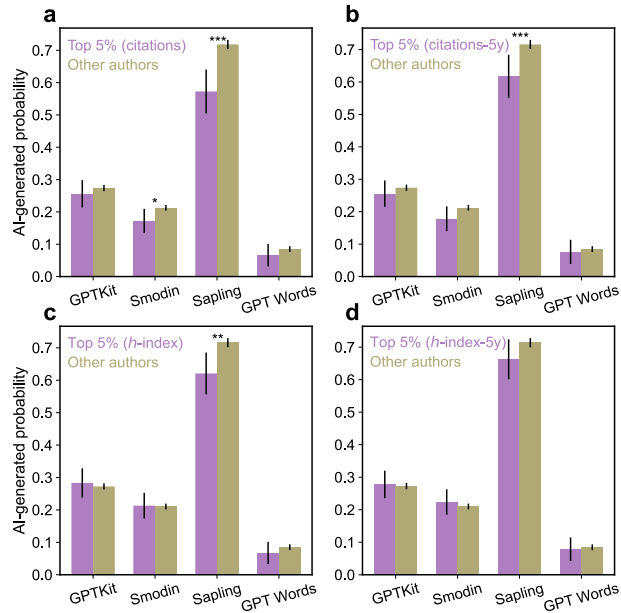

Figure 3. (a-d) Average AI-generated probabilities for authors ranked in the top 5% by (a) citation counts, (b) citation counts from the past five years, (c) $h$-index, and (d) $h$-index from the past five years. Error bars represent 95% confidence intervals.

Table 1. Spearman correlation coefficients between first author's academic performance and the four measurements of AI-generated text.

|  | Citations | Citations-5y | $h$-index | $h$-index-5y |
|---|---|---|---|---|
| GPTKit | -0.02 | -0.02 | -0.01 | -0.01 |
| Smodin | -0.02 | -0.02 | 0.01 | 0.01 |
| Sapling | -0.08*** | -0.07*** | -0.07*** | -0.06*** |
| GPT Words | -0.01 | -0.01 | 0.01 | 0.01 |

*3.4 Scientists who have tried AIGC have higher probability to use it in future academic writing*
Finally, we studied how previous adoption of AIGC in academic writing correlates with the AI-generated probability of future abstracts. We compared the AI-generated probabilities of abstracts from authors who have 0, 1, or more than 1 records in our dataset with an AI-generated probability over 0.9. As observed in Figure 4, the AI-generated probabilities increase as authors publish more papers with high probabilities of AI-generated content. This positive correlation suggests a common pattern of recurring use of AIGC in academic writing, with scientists likely becoming reliant on, or even accustomed to, AIGC tools due to their effectiveness and convenience.

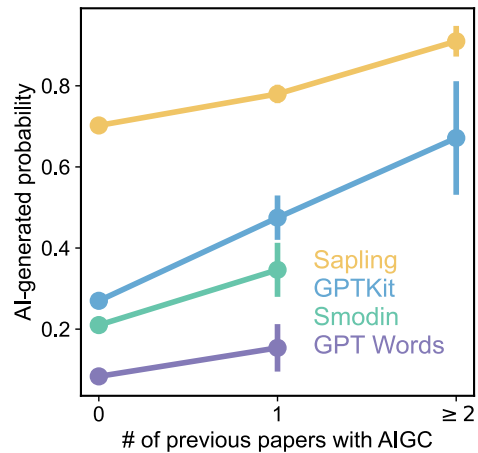

Figure 4. Average AI-generated probability for abstracts from first authors who have published 0, 1, or >1 papers with an AI-generated probability over 0.9. Error bars represent 95% confidence intervals. Results for Smodin and GPT Words on ">1" papers are not presented because the very limited number of data samples.

## 4. Discussion
This study quantitatively explored the use of AIGC in academic writing by examining preprints submitted to the arXiv website in 2023. We found that AIGC tools have made a significant influence. Since the release of ChatGPT, AI-generated probabilities have gradually increased, with nearly 10% of abstracts now displaying a high likelihood of being AI-generated. Further analysis revealed the correlation between AI-generated probability and three author-level factors: native languages, academic performance, and history of AIGC use. We found that: (1) scientists from English-speaking countries are less likely to use AIGC for writing assistance, whereas those from countries with a greater linguistic distance from English are more likely to use AIGC; (2) there is little correlation between the author's academic performance and the likelihood of using AIGC; and (3) authors who have published papers with high AI-generated probabilities exhibit higher possibilities to use AIGC tools in the future.

Some people hold a negative view of using AIGC in academic writing, considering it a form of academic misconduct. Our research reveals that AIGC tools do not reflect lower academic standards but, instead, assist non-native English speakers in reducing the complexities of academic writing, thereby facilitating better expression of their ideas. This can help mitigate the integration costs imposed by English dominance in academia, promoting equity and collaborative progress in science (Kozlowski et al., 2022). We are currently living in an era where AIGC profoundly influences scientific research. Many publishers now have allowed the use of AIGC for manuscript polishing, provided its use is disclosed (Picazo-Sanchez & Ortiz-Martin, 2024). Nonetheless, improving English proficiency and academic expression remains beneficial for scientists as AIGC-generated text still varies from human language patterns. We strongly advise against using unmodified AIGC outputs directly or for generating new viewpoints and sections. Additionally, as contents of scientific publications might contain critical information and confidential content, the ethical and privacy concerns associated with using AIGC tools need more attention.

There are still several limitations in this study, and we hope to continue our research to deepen our understanding of this topic. Firstly, our detection of AI-generated text does not necessarily reflect actual use of AIGC tools. AI-generated text might also be manually modified and/or polished by scientists, and instances of using unaltered AIGC outputs are less common. Our approach primarily measures the writing style of abstracts and does not assume that papers with high AI-generated probabilities are solely the result of AIGC assistance. Therefore, it is crucial to integrate our findings with other existing research, such as surveys, and to avoid overinterpreting our results. Secondly, our results are based solely on the field of large language models and cannot be directly generalized to other fields. We believe that it would be beneficial to expand our dataset to cover more disciplines, and conducting full-text analysis by parsing the PDF versions of these papers could produce insightful results. Finally, our study of the relationship between author profiles and AIGC use is currently limited to correlation analysis and does not imply causality. We hope that we employ more variables and causal inference methods to improve it in the future.

**Open science practices**
The data used is fully open available. See details in the footnotes when each dataset was firstly mentioned in this manuscript.

**Acknowledgments**
The authors are grateful to all members at the Knowledge Discovery Lab, Peking University. The language of part of this paper has been polished by ChatGPT.

**Author contributions**
Jialin Liu: Conceptualization, data curation, formal analysis, methodology, visualization, and writing-original draft.
Yi Bu: Conceptualization, formal analysis, funding acquisition, methodology, project administration, supervision, and writing-review and editing.

**Competing interests**
The authors declare no competing interests.

**Funding information**
This work was supported by the National Natural Science Foundation of China (#72104007 and #72174016).